\documentclass[journal]{IEEEtran}
\usepackage{cite,graphicx,amsmath,amssymb}
\usepackage{amsfonts,amssymb}
\usepackage{subfigure}
\usepackage{fancyhdr}
\usepackage{mdwmath}
\usepackage{mdwtab}
\usepackage{balance}
\usepackage{xcolor}
\usepackage{bm}
\usepackage{amsthm}
\usepackage{multirow}
\usepackage{flafter}
\usepackage{mathrsfs}

\usepackage{textcomp}
\makeatletter
\newif\if@restonecol
\makeatother

\usepackage[linesnumbered,ruled,vlined]{algorithm2e}
\usepackage{algpseudocode}
\usepackage{amsmath}

\usepackage{url}
\usepackage[
top    = 1.70cm,
bottom = 0.82 in,
left   = 0.63 in,
right  = 0.63 in]{geometry}

\theoremstyle{definition}
\newtheorem{definition}{Definition}

\newtheorem{theorem}{Theorem}
\newtheorem{remark}{Remark}
\newtheorem{corollary}{Corollary}

\renewcommand{\algorithmicrequire}{\textbf{Initialization}}   %

\begin{document}

\title{Resource Allocation in Wireless Powered IoT Networks}
\author{
\IEEEauthorblockN{Xiaolan~Liu,~Zhijin~Qin,~Yue~Gao,~Julie~A.~McCann
 }
%
%
\thanks{X. Liu, Z. Qin and Y. Gao are with the Department of Electronic Engineering and Computer Science, Queen Mary University of London, London E1 4NS, UK (email: \{xiaolan.liu, z.qin, yue.gao\}@qmul.ac.uk).

Z. Qin is also with Imperial College London as a Honorary Research Fellow.

J. A. McCann is now with Imperial College London, SW7 2AZ, London, UK (email:  j.mccann@imperial.ac.uk).
}
 }

\maketitle
\begin{abstract}
In this paper, efficient resource allocation for the uplink transmission of wireless powered IoT networks is investigated. We adopt LoRa technology as an example in the IoT network, but this work is still suitable for other communication technologies.
Allocating limited resources, like spectrum and energy resources, among a massive number of users faces critical challenges.
We consider grouping wireless powered IoT users into available channels first and then investigate power allocation for users grouped in the same channel to improve the network throughput. Specifically, the user grouping problem is formulated as a many to one matching game. It is achieved by considering IoT users and channels as selfish players which belong to two disjoint sets. Both selfish players focus on maximizing their own utilities. Then we propose an efficient channel allocation algorithm (ECAA) with low complexity for user grouping. Additionally, a Markov Decision Process (MDP) is used to model unpredictable energy arrival and channel conditions uncertainty at each user, and a power allocation algorithm is proposed to maximize the accumulative network throughput over a finite-horizon of time slots. By doing so, we can distribute the channel access and dynamic power allocation local to IoT users.
Numerical results demonstrate that our proposed ECAA algorithm achieves near-optimal performance and is superior to random channel assignment, but has much lower computational complexity. Moreover, simulations show that the distributed power allocation policy for each user is obtained with better performance than a centralized offline scheme.
\end{abstract}


\section{Introduction}
Predictions of the Internet of Things (IoT) describe an accumulation of potentially large number of interconnected IoT users to implement applications found in smart cities, smart agriculture and Industry 4.0. Therefore, allocating limited resources, such as spectrum and power, to a massive number of IoT users is a crucial and yet complex challenge. Traditionally, IoT users were battery powered which limits their operation time. Furthermore, many of these users are located in inaccessible spaces, remote areas or in hostile and toxic environments\cite{AD6_huang2014optimal}. For these reasons, there has been a strong movement towards the use of energy harvesting (EH) technologies\cite{AD7_lee2018multi}, where the device obtains energy from the multitude of renewable energy resources, like solar, wind, radio frequency (RF) signals and etc.. Specifically, the latter EH technique has been gaining more attention due to the proliferation of transmitters that already exist in our environment\cite{AD8_lu2015wireless}.

In many IoT applications, such as environment monitoring, although there is a large number of IoT users, each user only generates a relatively small amount of data. This enables the gateway to collect data from a massive number of devices, some communication technologies have been investigated to connect those devices, a cognitive radio enabled Time Division Long Term Evolution (TD-LTE) test-bed has been proposed to dynamically access spectrum over TV white space in\cite{1_gao2016scalable}. A relatively new class of communication protocols, described as low-power wide-area (LPWA) networks, are coming to the fore as they consume lower power in the IoT device while their transmission scope covers geographically larger area. Moreover, LPWA offers a trade-off among data transmission rate, network coverage and power consumption, to meet a large variety of requirements on IoT applications\cite{Xiong_magazine:2015,Mischa_JSAC:2016,Raza:Survey2017,zhijin2018,new_qin2017resource}. LPWA systems achieve this balance of power and distance at the cost of low data transmission rate which makes it more appropriate for latency-tolerant IoT applications with smaller data requirements.

The existing LPWA technologies contain LoRa\cite{LoRa_specification}, NB-IoT \cite{Sigfox} and Sigfox\cite{3gpp_release13}. Specially, LoRa is considered as one of the most potentially LPWA techniques, and has drawn much attention from both academics and industry \cite{Kartakis:2016,Bor:2016:LLW,Bor:EWSN:2016,Mo:IoT2016,Georgiou:2017}. LoRa technology achieves a big success mainly because it adopts chirp spread spectrum (CSS) modulation. The CSS technology contributes to flexible long range communications with low power consumption through using different spreading factors (SF)\cite{AD12_adelantado2017understanding}. Specifically, in LoRaWANs, the gateway server configures different SFs for users and there exists a mechanism to allow adaptive data rates (ADR). The server adjusts the SF and increases/reduces the transmit powers of the users according to its required signal-to-noise ratio (SNR) to optimize different metrics of the network including data transmission rate, airtime, and power consumption.

Battery-powered IoT devices suffering from finite energy, battery replacement and damage penalty cost, have been quantified specifically for devices running in LoRa\cite{addd8_imran2018energy}, which demonstrates that EH technique reduces the overall cost of the IoT solution to one-fifth of battery based systems. There are a multitude of EH resources such as solar, wind, electromagnetic signals and etc.. Indeed, EH technology has been extended to power the LoRa gateways to untether them in the same way as the IoT devices\cite{AD9_sherazi2018renewable}. Among all energy resources, RF signals are gaining more attention as they can be harvested from dedicated transmitters and are a good fit to the demand for far-field wireless power transfer (WPT). A RF-EH prototype has been designed and implemented in\cite{new5_muncuk2018multi}, which demonstrated the feasibility of a communication system which considers ambient RF signals as its only RF
resource to power battery-free devices.

\subsection{Related Work}
Wireless powered communication networks (WPCNs) have been proposed in IoT networks where the power beacons are regarded as the RF resource which can be harvested by the IoT devices and used to transmit their data \cite{new1_bi2016wireless,new2_ju2014throughput,new3_liu2014multi,new4_asiedu2018joint}. \cite{new2_ju2014throughput} has proposed a harvest-then-transmit protocol, in which a hybrid access point provides the users with wireless energy in the downlink and then users send their data in the uplink by using the harvested energy before. Here the sum of network throughput maximization problem was studied by exploring the optimal time allocation for the communication links and data transmission. In \cite{new3_liu2014multi}, the max-min problem of network throughput has been analyzed in WPCNs, in which users harvest energy from a multi-antenna access point (AP) and send their data to the same AP in the uplink.
The downlink-uplink time allocation, downlink energy beamforming and uplink transmit powers were jointly optimized using the non-negative matrix theory. In \cite{new4_asiedu2018joint}, the sum throughput rate maximization problem was analyzed where mobile stations with EH capabilities communicate with a full duplex multi-antenna AP. A joint optimization problem has been formulated with optimizing channel assignment, time allocation, and transmit power allocation simultaneously. \cite{AD7_lee2018multi} has proposed a novel floating device, integrating multi-source EH technologies and communication systems together to achieve long range communications with LoRa.

The performance of LPWA networks has been analyzed in \cite{Zhijin:ICC:2017}, which also analyzed the influence of interferences caused by LoRa and other LPWA technologies sharing the same frequency on the scalability of LoRa networks but this did not consider EH for devices. Renewable energy resources mostly belong to unpredictable behaviours which cause varieties amount of harvested energy. Power allocation based on an unknown amount of harvested energy becomes a challenging problem, in which the transmitter decides its transmit power according to the current available power stored in the battery only. Therefore, we use Markov Decision Process (MDP) to model the energy arrival process.
The solutions of MDP models are named as policies (e.g., transmit power policies), and those policies denote a series of best actions taken at each moment according to the status of the ambient environment. In addition, in MDP, we define the best actions as the optimal responses for maximizing the cumulative reward over the entire lifespan of a node.

In \cite{Ad2_loodaricheh2016qos}, an online resource allocation algorithm has been proposed, which adopted MDP to model the stochastic environment. It considered a system which contained a single source node and multiple destination nodes, then MDP was used to model the harvested energy and channel gain  guaranteeing the quality-of-service (QoS) of users. To address the throughput maximization problem, \cite{AD11_zenaidi2017performance} has introduced a first-order MDP to model the stochastic energy arrival, and then an optimal policy of transmit power allocation was obtained through solving dynamic programming. An expected data transmission maximization problem has been investigated in \cite{AD5_mao2014joint} with the constraints of energy harvesting rate, available battery energy, buffer queue and channel gain. An infinite-horizon discounted MDP was proposed to formulate this problem and it was solved by an optimal energy allocation algorithm. In \cite{addd9_sakulkar2018online}, online learning algorithms were adopted to exploit the MDP problem in which the rewards expectation and the state-action pairs were unknown, and then the algorithms were used to learn these rewards and propose their own policies over time. An MDP-based transmission policy for transmission duty cycle (TDC)-constrained networks has been derived to alleviate the TDC limitations in LPWA networks \cite{addd10_sandoval2018optimal}, and the feasibility of this policy has been confirmed via LoRa and Sigfox.

Although many research works have focused on the theory of decision making (e.g., MDP) in LPWA networks \cite{Ad2_loodaricheh2016qos,AD11_zenaidi2017performance,AD5_mao2014joint,addd9_sakulkar2018online,addd10_sandoval2018optimal}, few of them have considered efficient low resource allocation caused by a massive number of devices coupled limited battery life. Specially, even though LoRaWANs adopt ADR schemes, each LoRa device updates its transmit power by 3 dBm over each step, but it is inefficient for resource allocation. This is inefficient since the device has to communicate with the server for only increasing a little bit power, and it will cause interference to the device that is using the same SF.
Therefore, it is necessary to design a more efficient resource allocation method for mitigating channel access conflicts and improving resource allocation efficiency with the increasing number of devices. Moreover, an optimal scheme is necessary for devices to allocate the transmit power by themselves and to avoid the interference of using the same SF. None of this work concerns EH scenarios.

We investigate uplink transmission for wireless powered IoT networks with wireless powered devices to make the network energy self-sustainable in this paper.
As far as we know, this paper firstly explores the dual channel allocation and dynamic power allocation problem for wireless powered IoT networks.

The major contributions in this paper are presented as:
\begin{enumerate}
  \item We formulate resource allocation in wireless powered IoT networks into a joint optimization problem that optimizes channel allocation and dynamic power allocation, where MDP is adopted to model the harvested energy and channel conditions uncertainty. To make it traceable, the problem is decoupled into two phases: i) allocating IoT users to available channels; ii) optimizing transmit power allocation of IoT users assigned to the same channel within the same time frame.
  \item In the first phase, we propose an efficient channel allocation algorithm, called ECAA, to assign channels to users, which is achieved by enabling users to self-match with the proper channels based on matching theory.
  \item Within each channel, a MDP-based optimal power allocation algorithm is proposed to obtain the optimal policy of power allocation for each user over finite-horizon time slots.
  \item Numerical results demonstrate that our proposed ECAA achieves near-optimal performance while this algorithm has much lower computational complexity than the brute force exhaustive-search approach. Moreover, the optimal power allocation policy is obtained with higher throughput performance than an alternative offline scheme.
\end{enumerate}

We organize the rest of this paper as follows. In Section II, we build up the system model and formulate the resource allocation problem in wireless powered IoT networks. In Section III, the proposed ECAA is illustrated to enable users self-matching with available channels. Then the proposed MDP-based power allocation algorithm is described in Section IV. Simulations are performed in Section V, finally we give the conclusions in Section VI.

\section{System Model and Problem Formulation}
\begin{figure}[t!]
    \begin{center}
        \includegraphics[width=3.7in]{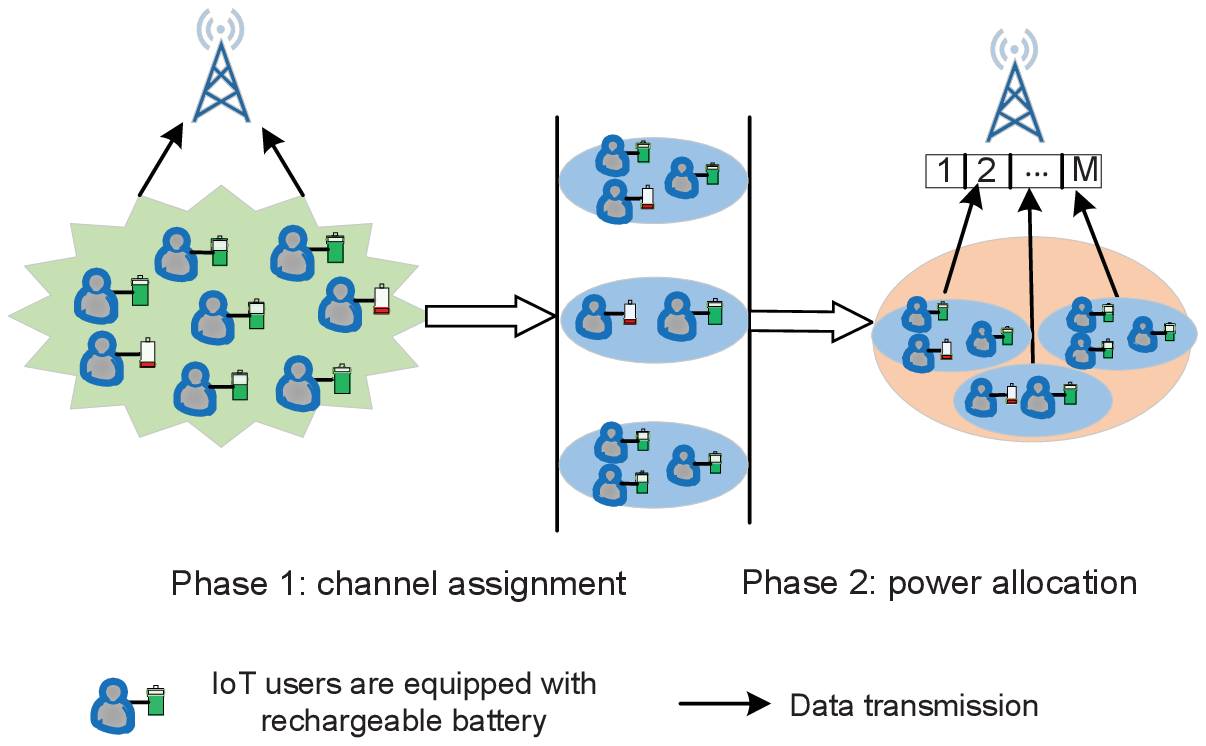}\\
        \caption{\small{The proposed framework for wireless powered IoT networks.}}
        \label{system_model}
    \end{center}
\end{figure}
In this paper, we focus on the uplink transmission of an IoT network where the gateway owns $M$ channels to be accessed. We assume each user is equipped with a RF energy harvester and it has a finite-capacity rechargeable battery.
The user is assumed to harvest energy from ambient environment, such as wireless power beam or broadcasting TV signals, as shown in Fig.~\ref{system_model}.
Each user wakes up for data transmission when the available power in the battery is higher than the pre-defined threshold. We define the number of active users as $N$, and the users  are located uniformly in a circle covered by the network. The channel set and the IoT user set are indicated as $\mathcal{L}=\left\{ {{{L}_1},\ldots ,{{L}_m}, \ldots ,{{L}_M}} \right\}$ and $\mathcal{U}=\left\{ {{U_1},\ldots ,{U_n}, \ldots ,{U_N}} \right\}$, respectively. We denote the bandwidth of any channel, ${L}_m$, as $B_m$~Hz.
\begin{figure}[t]
    \begin{center}
        \includegraphics[width=2.5in]{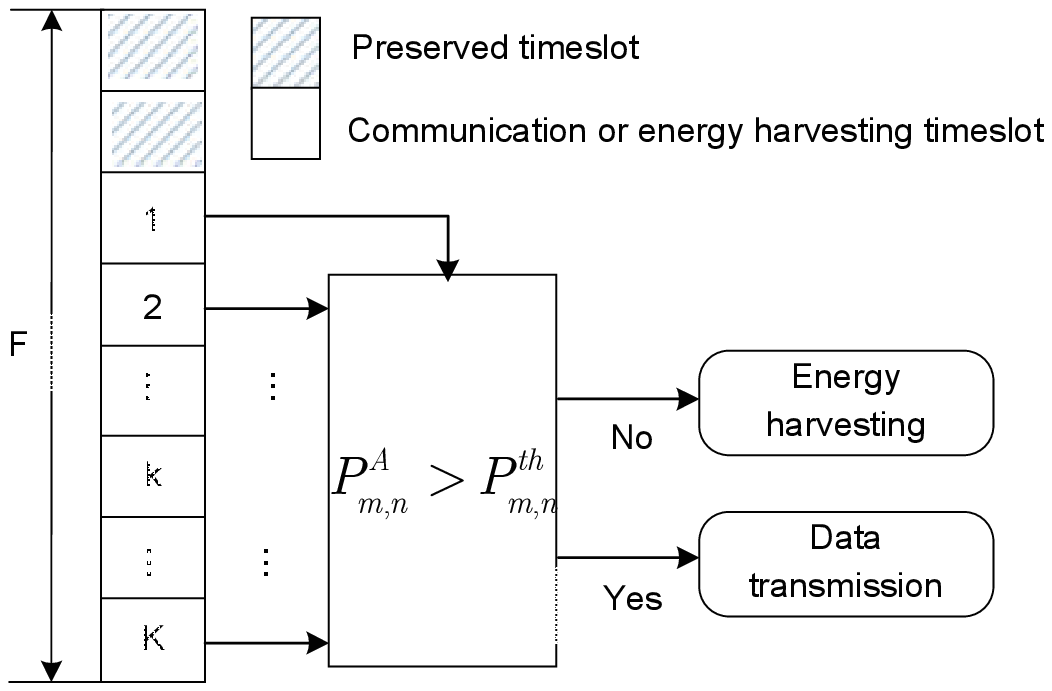}
        \caption{\small{The proposed frame model for each wireless powered IoT user.}}
        \label{timeslot_model}
    \end{center}
\end{figure}

Although our scheme is general, we use LoRa as an illustrative example. In this case, both the gateway and IoT user are assumed to be equipped with a single antenna as defined in ~\cite{LoRa_specification}. We partition each frame $F$ into different time slots which are used for channel allocation, energy harvesting as well as data transmission as shown in Fig.~\ref{timeslot_model}. The IoT network is working as frame-based structure over time, and all the users adopt the harvest-then-transmit protocol. Specifically, we assume each frame contains $K>1$ slots, indexing from 1 to $K$. In each frame, assuming that $t$ indicates the number of time slots used to transmit data, and except for the first few time slots reserving for channel allocation, the remaining slots are assigned to users to transmit data.

From Fig.~\ref{timeslot_model}, at the beginning of each time slot,
comparing the battery available power to the required power for data transmission, if $P_{m,n}^A(k) > P_{m,n}^{th}(k)$, this time slot is assigned to data transmission,  otherwise it is used to harvest energy.
The harvested power of each user is randomly over time, then we use $P_{m,n}^H(k)$ to denote the amount of power harvested by user $n$ in channel $m$ over time slot $k$. Let $\bm{{P^H}(k)}: = [P_1^H(k),...P_q^H(k)...,P_Q^H(k)]$ be the harvested power state at $k$, which has $Q$ possible values.
Denote the battery capacity as $P_{max}^A$ with the transmit power of $U_n$ satisfying $P_{m,n}(k) \le P_{max}^A < \infty $, then the power available at the beginning of each time slot is presented as
\begin{align}\label{energy_dynamic}
P_{m,n}^{A}(k+1) = \min (P_{m,n}^{A}(k) - P_{m,n}(k) + P_{m,n}^{H}(k),\;P_{max}^A).
\end{align}
Since we discretize time slots with data transmission and energy harvesting as shown in Fig.~\ref{timeslot_model}, the update of available power at each time slot as given by (\ref{energy_dynamic}) is reformulated into two cases depending on the usage of the time slot. When the time slot is used for harvesting energy, the transmit power is $P_{m,n}(k)=0$, (\ref{energy_dynamic}) can be expressed as
\begin{align}\label{energy_dynamic1}
P_{m,n}^{A}(k+1) = \min (P_{m,n}^{A}(k) + P_{m,n}^{H}(k),\;P_{max}^A).
\end{align}
But when it is used for data transmission, the harvested power becomes $P_{m,n}^{H}(k)=0$, (\ref{energy_dynamic}) can be expressed as
\begin{align}\label{energy_dynamic2}
P_{m,n}^{A}(k+1) = \min (P_{m,n}^{A}(k) - P_{m,n}(k),\;P_{max}^A).
\end{align}
We assume the channel from the IoT user to the gateway $U_n$ suffers from Rayleigh fading. Thus, the channel gain $g_{m,n}$ over a time slot $k$ is given by
\begin{align}\label{channel_gain}
{g_{m,n}(k)} = {h_{m,n}(k)}{\eta _m}{\left\| {{d_{m,n}}} \right\|^{ - a}},
\end{align}
where $h_{m,n}(k)$ presents the small-scale fading of ${L}_m$ which is measured at $U_n$, $\eta_m$ is a path loss parameter of communication link in ${L}_m$. The distance between the gateway and IoT user, $U_n$, is denoted by $d_{m,n}$,
and it will not change over different time slots because all the users are assumed static in the network. The path loss exponent $a$ is decided by the carrier frequencies and environmental conditions, therefore, the signal received at the gateway over ${L}_m$ is presented as
\begin{align}
{y_m}(k) = \sum\limits_{n = 1}^N {{\alpha _{m,n}}{P_{m,n}(k)}{g_{m,n}(k)}}  + \sigma _m^2,
\end{align}
where the noise is assumed as additive white Gaussian noise (AWGN) with its power $\sigma _m^2$, $P_{m,n}(k)$ presents the transmit power of $U_n$ in the uplink when transmitting data over ${L}_m$ in time slot $k$. We use $\alpha_{m,n}$ to indicate whether ${L}_m$ is assigned to user $U_n$, which can be defined as
\begin{align}\label{alpha}
{\alpha _{m,n}} = \left\{ {\begin{array}{*{20}{c}}
1,&\text{$U_n$\;occupies\;$L_m$,}\\
0,&\text{otherwise.}
\end{array}} \right.
\end{align}
We denote $M_m$ as the number of users assigned to ${L}_m$, i.e., ${M_m} = \sum\limits_{n = 1}^N {{\alpha _{m,n}}} $. We assume that the users assigned into the same channel use different SFs to transmit data in the same time frame. Since signals with different SFs are implemented by orthogonal pseudo-random codes, the interferences among them can be ignored\cite{AD12_adelantado2017understanding}. So the uplink SNR for $U_n$ transmitting over ${L}_m$ in time slot $k$ can then be calculated as
\begin{align}\label{SINR}
{\gamma _{m,n}(k)} = \frac{{{P_{m,n}(k)}{g_{m,n}(k)}}}{{ \sigma _m^2}},\;\forall \;m,\;n,\;k.
\end{align}
Then for giving any user, $U_n$, its achievable accumulative data rate in ${L}_m$ over any time frame can be given by
\begin{equation}\label{data_rate}
\begin{array}{l}
R_{m,n} = \sum\limits_{k = 1}^{t}R_{m,n}(k)\\
\;\;\;\;\;\;\;\;\;=\sum\limits_{k = 1}^{t}{B_m}{\log _2}\left( {1 + {\gamma _{m,n}(k)}} \right),\;\forall \;m,\;n.
\end{array}
\end{equation}
Our objective is to maximize the network throughput with increasing resource allocation efficiency for IoT networks under dynamic energy harvesting constraints. Then we can formulate the problem as
\begin{subequations}\label{formulated_pro}
\begin{align}
\left( \textbf{{P1}} \right)
&\mathop {\max }\limits_{{\alpha _{m,n}},\;{P_{m,n}}(k),\;t} \sum\limits_{n = 1}^N {\alpha _{m,n}}{R_{m,n}}(t,\;P_{m,n}(k)),\\
\text{s.t.}\;\;
&C_1:\;0 \le{ {P_{m,n}(k)} } \le P_{m,n}^{A}(k)\;\;\forall \;m,\;n,\;k,\label{C2} \\
&C_2:\;t \in \{ 1,...,K \} ,\label{C3}\\
&C_3:\;{\alpha _{m,n}} \in \left\{ {0,1} \right\},\;\forall\;m,\;n,\label{C4}\\
&C_4:\;\sum\limits_m {{\alpha _{m,n}}}  \le D,\;\forall\; n,\label{C5}\\
&C_5:\;\sum\limits_n {{\alpha _{m,n}}}  \le 1,\;\forall\; m.\label{C6}
\end{align}
\end{subequations}
where $C_1$ denotes value range of the transmit power for any user $U_n$ over $L_m$ in any time slot, $P_{m,n}^{A}(k)$ is the battery power available at the beginning of time slot $k$, and it is decided by (\ref{energy_dynamic}). In $C_2$, $t$ is the number of time slots allocated to transmit data in each frame.
In $C_3$, $\alpha_{m,n}$ is let to be either 0 or 1. There are at most $D$ users assigned to one channel, which is shown in~$C_4$, and~$C_5$ is used to restrict that a user at most can only be accessed to one channel, $L_m$.

It is noted that the problem \textbf{(P1)} involves integer programming as shown in $C_2$, the binary constraints as well as stochastic constraint in the objective function, so it is obviously a non-convex problem. As a result, there is no efficiently computational approach to solve the optimization problem \textbf{(P1)}. Thus, from Fig.~\ref{system_model}, we are proposing to decompose this optimization problem into two phases and solve them separately. In the channel allocation phase,
all the IoT users are self-matched with available channels. An MDP-based power allocation algorithm over finite-horizon is proposed for the wireless powered users assigned with the same channel in the second phase.

\section{Efficient Channel Allocation Algorithm}
Assuming each user transmits data to the gateway using the same transmit power, the channel allocation problem can be presented as
\begin{align}\label{channel_allocation}
\left( \textbf{{P2}} \right)\;\;
\mathop {\max }\limits_{\alpha_{m,n}}\; {R_{m,n}^{uti}},\;
\text{subject to $C_3$, $C_4$, and $C_5$},
\end{align}
where the $R_{m,n}^{uti}$ represents the utility of $U_n$ and ${L}_m$. To consider the fairness of all the users, \textbf{(P2)} is transferred to a max-min problem.
So the utility of $U_n$, ${R_{U_n}^{uti}}$, is defined as the minimal transmission rate of user, $U_n$, among those channels it occupies $\mathcal{J}_n$ at any time slot $k$, which is expressed as
\begin{equation}\label{user_n_data_rate}
\begin{array}{l}
{R_{U_n}^{uti}} = \min\left(  {{B_m}{{\log }_2}(1 + \gamma _{m,j}(k)) }\right),
\forall\; j\in \mathcal{J}_n.
\end{array}
\end{equation}
Moreover, the utility of ${L}_m$, denotes the minimal transmission rate from the set of users, $\mathcal{I}_m$, which share the same channel ${L}_m$ at any time slot $k$, can be described as
\begin{equation}\label{channel_m_data_rate}
\begin{array}{l}
{R_{L_m}^{uti}} = \min\left(  {{B_m}{{\log }_2}(1 + \gamma _{m,i}(k)) }\right),
\;\forall\; i \in \mathcal{I}_m.
\end{array}
\end{equation}

Note that (\textbf{P2}) is NP-hard. We propose an efficient channel allocation algorithm, named ECAA, with low complexity based on matching theory to solve it. The proposed ECAA reduces the probability of re-transmission since channel conflicts between any two users are eliminated, which extends the battery life of users. In our proposed algorithm, the IoT user  set, $\mathcal{U}$, and channel set, $\mathcal{L}$, are considered as two disjoint sets of selfish players which only focus on maximizing their own utilities. Moreover, assuming the channel state information (CSI) is known \footnote{The CSI is recorded at the gateway after it receives the requests from users and it is broadcasted to all the users in the next downlink data package.}, that is, each user is aware of the CSI of other users. We will discuss more details of the proposed algorithm in Section~\ref{ECAA}.

\subsection{Many-to-One Matching}
The basic knowledge of the many-to-one matching model is introduced to address channel allocation problem~\cite{Han:SPM} in this section.

\subsubsection{Matching pair}
In this paper, a matching can be considered as matching channels in the channel set $\mathcal{L}$ with users in the user set $\mathcal{U}$, the formal definition is given in the following.
\theoremstyle{definition}
\begin{definition}\label{matching_definition}
By giving two disjoint sets, the channel set $\mathcal{L}$ and the user set $\mathcal{U}$, a many-to-one matching $\Phi $ is defined to map the channel set $\mathcal{L} \cup  \mathcal{U}$ to the user set including all subsets of $\mathcal{L} \cup  \mathcal{U}$ so that for each ${L}_m\in \mathcal{L}$ and $U_n \in \mathcal{U}$:
\begin{enumerate}
  \item $\Phi  \left({L}_m\right)\subseteq \mathcal{U}$ ;
  \item $\Phi  \left({U}_n\right)\subseteq \mathcal{L}$;
  \item $\left| {\Phi  \left( {{U_n}} \right)} \right| \le 1$;
  \item $\left| {\Phi  \left( {L_m} \right)} \right| \le D$;
  \item ${L_m} \in \Phi  \left( {{U_n}} \right) \Leftrightarrow {U_n} \in \Phi  \left( {{L_m}} \right)$.
\end{enumerate}
The meanings of the aforementioned 5 conditions are as follows:
1) lets each channel, $L_m$, match with a subset of users, $\mathcal{U}$;
2) lets each user, $U_n$, match with a subset of channels, $\mathcal{L}$.
3) implies that each user at most can access one channel at a time;
4) restricts the number of users assigned to one channel, which cannot exceed $D$.
5) is the defined matching pair.
\end{definition}

\begin{remark}\label{peer_effects}
From $\textbf{Definition1}$, the formulated matching game can be considered as a many-to-one problem.
\begin{proof}
A many-to-one matching game is formulated for our channel allocation problem, this is because each user only matches with no more than one channel at a time, while each channel can be accessed by more than one users. Each player is concentrated on maximizing its own utility~\cite{Bodine-Baron2011}.
\end{proof}
\end{remark}

\subsubsection{Preference relations}
In this part, a preference relation, $\succ $, is defined to illustrate competition behaviors and decision making for both users and channels. Particularly, given any user, $U_n \in \mathcal{U}$, its preference $\succ_{U_n}$ between any two channels, ${L}_m \in \mathcal{L}$ and ${L}_{m'} \in \mathcal{L}$ with $m \ne m'$, is presented as
\begin{align}\label{preference}
\left( {{L_m},\Phi  } \right){ \succ _{{U_n}}}\left( {{L_{m'}},\Phi  '} \right) \Leftrightarrow {R_{m,n}}\left( \Phi   \right) > {R_{m',n}}\left( {\Phi  '} \right),
\end{align}
where ${L_m} \in \Phi  \left( {{U_n}} \right)$, ${L_{m'}} \in \Phi ' \left( {{U_n}} \right)$. This definition implies that user $U_n$ prefers ${L}_m$ in $\Phi $ to ${L}_{m'}$ in $\Phi '$ if ${L}_m$ can provide higher transmission rate than ${L}_m'$. Likewise, given any channel ${L}_m$, its preference $\succ_{{L}_m}$ between any two set of users, $\mathcal{S_U}\in \mathcal{U}$ and $\mathcal{S_U'}\in \mathcal{U}$, is derived as
\begin{align}\label{preference1}
\left( {\mathcal{S_U},\Phi  } \right){ \succ _{{{L}_m}}}\left( {\mathcal{S_U'},\Phi '} \right) \Leftrightarrow {R_{m,n}}\left( \Phi   \right) > {R_{m,n'}}\left( {\Phi  '} \right),
\end{align}
where $ \mathcal{S_U} \in \Phi  \left( {{{L_m}}} \right)$ and $ {\mathcal{S_U'}}\in \Phi ' \left( {{L_{m}}} \right)$.

\subsubsection{Swap matching}
From the matching game, we define the swapping behaviours of players as that each two players are supposed to swap their matching but won't change any other players' assignment. Therefore, we give the detailed concept of \emph{swap-matching} to better explain the interdependency of players' preference.
\begin{definition}
Given a matching $\Phi $ with ${L_m} \in \Phi  \left( {{U_n}} \right)$, ${L_{m'}} \in \Phi  \left( {{U_{n'}}} \right)$, ${L_m} \notin \Phi  \left( {{U_{n'}}} \right)$, and ${L_{m'}} \notin \Phi  \left( {{U_n}} \right)$, a swap matching $\Phi  ' = \left\{ {\Phi  \backslash \left\{ {\left( {{U_n},{L_m}} \right),\left( {{U_{n'}},{L_{m'}}} \right)} \right\}} \right\} \cup \left\{ {\left( {{U_n},{L_{m'}}} \right),\left( {{U_{n'}},{L_m}} \right)} \right\}$ is defined by ${L_m} \in \Phi  '\left( {{U_{n'}}} \right)$, ${L_{m'}} \in \Phi  '\left( {{U_n}} \right)$, ${L_m} \notin \Phi  '\left( {{U_n}} \right)$, and ${L_{m'}} \notin \Phi  '\left( {{U_{n'}}} \right)$.
\end{definition}

It is noticed the swap matching defines a matching generated by a swap operation.
The swap-blocking pair is defined as follows.
\begin{definition}\label{D3}
Giving a user pair $\left( {{U_n},{U_{n'}}} \right)$ that are matched for a given matching $\Phi $, if there is ${L_m} \in \Phi  \left( {{U_n}} \right)$ and ${L_{m'}} \in \Phi  \left( {{U_{n'}}} \right)$ so that:
\begin{enumerate}
  \item $\forall i \in \left\{ {{U_n},{U_{n'}},{L_m},{L_{m'}}} \right\}$, ${R_i}\left( {\Phi  '} \right) \ge {R_i}\left( \Phi   \right)$ and
  \item $\exists i \in \left\{ {{U_n},{U_{n'}},{L_m},{L_{m'}}} \right\}$ such that ${R_i}\left( {\Phi  '} \right) > {R_i}\left( \Phi   \right)$,
\end{enumerate}
so the swap matching $\Phi '$ is defined, and we define $\left( {{U_n},{U_{n'}}} \right)$ as a swap-blocking pair in $\Phi $.
\end{definition}

\begin{algorithm}[t]
\caption{Efficient Channel Allocation Algorithm (ECAA)}
\label{matching_theory_LPWA}
\renewcommand{\algorithmicrequire}{\textbf{Initialization}}
\KwIn{ The initial matching $\Phi _0$ generated by Algorithm~\ref{IA}.}   

\While {The current matching is blocked by $\exists \left( {{U_n},{U_{n'}}} \right)$ }
{
    \For {$\forall {U_n} \in \mathcal{U}$}
    {
        \For {$\forall {U_{n'}} \in \left\{ {\mathcal{U}\backslash {U_n}} \right\}$ with ${{L}_m} \in \Phi  \left( {{U_n}} \right)$ and ${L_{m'}} \in \Phi  \left( {{U_{n'}}} \right)$}
        {

        \If {a swap-blocking pair $\left( {{U_n},{U_{n'}}} \right)$ appears and $C_2-C_4$ are satisfied}
        {
            ${U_n}$'s match $L_m$ is exchanged with ${U_{n'}}$'s match $L_{m'}$.

            Update $\Phi $.
        }

      }
   }
}
\Return the final matching $\Phi $.
\end{algorithm}

The success of a swap matching operation implies that the utility of a arbitrary player, i.e., $R_{U_n}$ or $R_{L_m}$ as shown in~\eqref{user_n_data_rate} and~\eqref{channel_m_data_rate}, will not decrease, moreover, it will at least increase the utility of one player. Note that both the users and the gateway can initialize the swap operation because the utilities of them are directly relevant to the data transmission rate.

\begin{definition}
If a matching $\Phi $ is not blocked by any swap-blocking pair, it is defined as two-sided exchange-stable (\textbf{\emph{2ES}}).
\end{definition}

\begin{algorithm}[t]
\caption{ Matching Initialization Algorithm}
\label{IA}
\KwIn{Given any user set $\mathcal{\Phi _{UM}}=\mathcal{U}$, $\alpha_{m,n}=0$,
initialize proposal indicator ${\beta}_{m,n}=0$, $\forall \;m,\;n$.}

Obtain the preference list of each user $\mathcal{PI}_{U_n} $, $\forall \;U_n \in \mathcal{U}$.

Obtain the preference list of each channel $\mathcal{PI}_{{L}_m}$, $\forall \;L_m \in \mathcal{L}$.

\While {$\mathcal{\Phi _{UM}}\ne \emptyset$}
{
    \For {$\forall {U_n} \in \mathcal{U}$}
    {
	${U_n} $ chooses its first preference that has not failed before.

 Update $\beta_{m,n}=1$ if  $U_n$ proposes to $L_m$.
    }
    \For{$\forall \; {L_m} \in \mathcal{L}$}
    {
    	\If {$\sum\limits_n {\left( {{\alpha _{m,n}} + {\beta _{m,n}}} \right)}  \le D$}
    {
		 ${L_m} $ accepts all proposals from users.
		}
\Else
{
 ${L_m} $ accepts proposals from the first preferred $D$ users.
	}

	 Update $\mathcal{\Phi _{UM}}$ by eliminating all the matched users $U_n$.
	Eliminate $L_m$ from $\mathcal{PI}_{U_n}$ if $\beta_{m,n}=1$.

	Update $\Phi _0$ with $\alpha_{m,n}=1$ for all the matched $U_n$.
    }
}
\If {there are vacant channels, $L_m$}
{
 Let $L_m$ match with its first preferred user.

Update $\Phi _0$.
}
\Return $\Phi _0$.
\end{algorithm}

\subsection{Proposed Efficient Channel Allocation Algorithm for IoT Networks}\label{ECAA}
In this section, an efficient channel allocation algorithm, named ECAA, based on matching theory is proposed, which has low complexity and is able to distribute channel allocation to users. The ECAA aims to look for a \textbf{\emph{2ES}} matching used for channel allocation after finishing a few swap operations. Furthermore, the battery life of users is extended since the ECAA has low computation complexity.

From Algorithm~1, the ECAA is proposed including initialization algorithm and swap matching algorithm. In the initialization step, we propose an initialization algorithm to generate the initial matching, $\Phi _0$, which is illustrated in Algorithm~\ref{IA}. we assume each user transmits data using the same transmit power. The preference list, i.e., the available channels, of each user is constructed based on the CSI. For example, given $U_n$, the channel with the best CSI, i.e., $m = \arg \mathop {\max }\limits_{\forall m} {{g_{m,n}}} $ is defined as the user's first preference. The preference list is initialized at the gateway for each user by calculating the distance between them and the gateway, that is, the highest preferred user is the closest one. This is because in LPWA networks, the achieved transmission rate is mainly affected by the large-scale fading. Therefore, each user chooses its first preferred channel from its preference list, and only the proposals of the first $D$ users in the preference list are accepted by each channel. Only when all the users are matched with a channel, this process can be stopped.
Specifically, if the channel is not matched with any user, we force it to match with its first preferred user, which is used to improve the minimal achievable data transmission rate. We return the initial matching $\Phi _0$ and consider it as an input for Algorithm~1.

In the swap-matching algorithm, users keep looking for swap-blocking pairs, while the utilities of both players are guaranteed not to decrease and at least the utility of one player, the user or the channel, will increase. The swap operation carries out if only one swap-blocking pair appears in the present matching game. The process of searching and swap operation is stopped when reaching the ultimate matching state. Then after Algorithm~1 returning the channel access decision, the preferred channel lists of active users are renewed.

A few theorems have been derived from our proposed ECAA, which are as follows.
\begin{theorem}\label{theorem1}
\textbf{Stability}: in the ECAA, the ultimate matching is a \textbf{\emph{2ES}} matching.
\begin{proof}
Please see Appendix A.
\end{proof}
\end{theorem}

\begin{theorem}\label{theorem2}
\textbf{Convergence}: after performing swap operations a finite number of times, ECAA is converged to the final matching, \textbf{\emph{2ES}}.
\begin{proof}
Please see Appendix B.
\end{proof}
\end{theorem}


\begin{theorem}\label{theorem3}
\textbf{Complexity}: there exists an upper bound of the computational complexity for the proposed ECAA, which is calculated by ${\rm O}\left(MN+ {\frac{1}{2}IDN\left( {M - 1} \right)} \right)$.
\begin{proof}
Please see Appendix C.
\end{proof}
\end{theorem}

It is easily noticed that the ECAA has much lower complexity than the brute force exhaustive-search approach in which the computation becomes more and more complicated, even is increasing exponentially, with the increasing number of active users, $N$.



\section{MDP-based Power Allocation Algorithm}
After grouping all the users by ECAA, they are allocated to one of the available channels. For those users allocated to the same channel, \textbf{(P1)} is simplified as

\begin{subequations}\label{reformulated}
\begin{align}
\left( \textbf{{P3}} \right)
&\mathop {\max }\limits_{{P_{m,n}}(k),\;t}\sum\limits_{n = 1}^N {R_{m,n}}(t,\;{P_{m,n}}(k)),\\
\text{s.t.}\;\;
&C_1:\;0 \le {P_{m,n}}(k) \le P_{m,n}^A(k)\;\;\forall m,\;n,\;k,\\
&C_2:\;t \in \{ 1,...K \}.
\end{align}
\end{subequations}

To increase the network throughput, a distributed scheme that optimizes the transmission rate of any user is developed.
We assume that data transmission and energy harvesting are performed separately in different time slots, that is, the transmit power $P_{m,n}(k)=0$ when the time slot ${k}\subseteq \mathcal{K}_h$ is allocated to harvest energy; similarly, the harvested power $P_{m,n}^H(k)=0$ when the time slot ${k}\subseteq \mathcal{K}_d$ is used to transmit data. Moreover, $ \mathcal{K}_h\cup \mathcal{K}_d=\mathcal{K}$, where $\mathcal{K}$ presents all the time slots in any time frame.
In this paper, the transmit power allocation is optimized over all the timeslots while the time slot used for energy harvesting can be considered as fixed power allocation with the transmit power $P_{m,n}(k_h)=0$.
A pre-defined power threshold $P_{m,n}^{th}$ is used to enable data transmission at the user. So the optimization problem is changed as
\begin{subequations}\label{reformulated1}
\begin{align}
\left( \textbf{{P3.1}} \right)
&\mathop {\max }\limits_{{P_{m,n}}(k)} {R_{m,n}}({P_{m,n}}(k)),\\
\text{s.t.}\;\;
&C_1:\;P_{m,n}^{th}\le {P_{m,n}}(k) \le P_{m,n}^A(k)\;\;\forall m,\;n,\;{k}\subseteq \mathcal{K}_d,\\
&C_2:\;{P_{m,n}}(k)=0 \;\;\forall m,\;n,\;{k}\subseteq \mathcal{K}_h.
\end{align}
\end{subequations}
where $C_1$ indicates the value range of the transmit power when the time slot is used for data transmission, and $C_2$ denotes the case that the time slot is used to harvest energy.
Since it is impossible to know all the knowledge of the future time slots, we assume that only some stochastic information of the harvested power $P_{m,n}^{H}(k)$ and the channel gain $g_{m,n}(k)$ for future time slots are available. Then we use finite-horizon MDP to model the joint random process of $P_{m,n}^{H}(k)$ and $g_{m,n}(k)$, and provide a MDP-based power allocation algorithm via dynamic programming\cite{Ad2_loodaricheh2016qos}.

Assuming the channel information is only available in current time slot, that is, $g_{m,n}(k)$ is known at time slot $k$. Note that the amount of harvested energy in time slot $k$ is unavailable until the end of the time slot, i.e., it will be known at time slot $k+1$. Thus, channel gain and energy arrival process are modelled by first-order Markov model. The EH process and the channel gain are assumed to be independent, then the transition probabilities of them are defined as $\textbf{P}_\textbf{r}(P_{m,n}^H(k)|P_{m,n}^H(k-1))$ and $\textbf{P}_\textbf{r}(g_{m,n}(k)|g_{m,n}(k-1))$. So the joint probability density function (PDF) is given by
\begin{equation}\label{joint_pdf}
\begin{array}{l}
\textbf{P}_\textbf{r}(P_{m,n}^H(k-1),\;g_{m,n}(k))\\
=\textbf{P}_\textbf{r}(P_{m,n}^H(k-1)|P_{m,n}^H(k-2))
\textbf{P}_\textbf{r}(g_{m,n}(k)|g_{m,n}(k-1)).
\end{array}
\end{equation}
As shown in (\ref{energy_dynamic}), the available power $P_{m,n}^A(k)$ in the time slot $k$ depends on its last state $P_{m,n}^A(k-1)$, which also can be modelled as a first order Markov model.

So the system states in the time slot $k$ is defined as
\begin{align}\label{state_space}
S_{m,n}(k)=(P_{m,n}^A(k), \;P_{m,n}^H(k-1), \;g_{m,n}(k)),
\end{align}
where $S_{m,n}(k)$ indicates the state space of user, $U_n$, assigned into channel, $L_m$, it consists of the available battery power, $P_{m,n}^A(k)$, and the channel gain, $g_{m,n}(k)$, in the current time slot $k$, as well as the harvested power, $P_{m,n}^H(k-1)$, in the previous time slot $k-1$. Based on the current state, $S_{m,n}(k)$, at time slot $k$, the user will decide to transmit data with power $P_{m,n}(k)$. That is an action taken at time slot $k$ from its feasible set $\pi_k$
\begin{equation}\label{action_set}
\begin{array}{l}
\pi_k=\{P_{m,n}^{th} \le P_{m,n}(k)\le P_{m,n}^A(k)\;and\; P_{m,n}(k)=0\},\\
\;\;\;\;\;\;\;\;\;\;\;\;\;\;\;\;\;\;\;\;\;\;\;\;\;\;\; \forall m,\;n,\;k.
\end{array}
\end{equation}
We consider a threshold power $P_{m,n}^{th}$ for each user, therefore, the action space is reduced with this measurement. In this case, some time slots become available for energy harvesting, and the action $P_{m,n}(k)=0$ is taken when this time slot is used to harvest energy.

\begin{algorithm}[t!]
\renewcommand{\algorithmicrequire}{\textbf{Inputs:}}
\caption{MDP-based Power Allocation Algorithm}
\label{optimal_online}
\KwIn{IoT users are assigned with one of the available channels according to ECAA.}

\textbf {\underline{Planning phase}}~\\
Set $k=K$, calculate $V_K(S_{m,n}(K)))$,
$\forall S_{m,n}(K)=\{P_{m,n}^A(K), P_{m,n}^H(K), g_{m,n}(K)\}$, using (\ref{bellman1})

\For {$k=K-1:1$}
{
Calculate $V_k(S_{m,n}(k))$ using (\ref{bellman2}), $\forall P_{m,n}^A(k), P_{m,n}^H(k)$
}
\textbf {\underline{Transmission phase}}~\\
Initializing {$P_{m,n}^A(k),\;g_{m,n}(k),\;P_{m,n}^H(k-1)$}\\

\For {$k=1:K$}
{
\If {$P_{m,n}^A(k)>P_{m,n}^{th}$}
{
Finding $\pi_k^*$ that maximize $V_k(S_{m,n}(k))$ from the planning phase, that is, $P_{m,n}^H(k)=0$
}
\Else
{
 The user harvests $P_{m,n}^H(k)$ amount of power, and the optimal policy in time slot $k$ $\pi_k^*=0$
}

IoT user $U_{m,n}$ transmits data to the gateway with the transmit power from $\pi_k^*$ at time slot $k$

Update the available power $P_{m,n}^A(k+1)$ in the battery through (\ref{energy_dynamic})
}
\Return $\pi_k^*$, $\forall k$
\end{algorithm}

We aim to obtain an optimal policy $\pi^*$ that maximizes the accumulative expected network throughput over a finite time frame which including $K$ time slots for each user. By giving an initial state, $S_{m,n}(0)$, the optimal value function can be calculated as
\begin{align}\label{optimal_value_0}
V^*=\mathop {\max }\limits_{\pi  \in \prod } \;\sum\limits_{k = 1}^K {{\mathbb E}\{ {R_{m,n}}(k)}
|{S_{m,n}(0)},\;\pi \},
\end{align}
where $\mathbb E$ indicates the statistical expectation of channel gain and the harvested energy.
Then our optimization problem is represented as
\begin{subequations}\label{reformulated3}
\begin{align}
\left( \textbf{{P3.2}} \right)
&\mathop {\max }\limits_{\pi_k} \;\;{\mathbb E}\{ [\sum\limits_{k = 1}^K {{R_{m,n}}(k)]|\textbf{P}_\textbf{r}} \}, \\
\text{s.t.}\;\;
&C_1:\;P_{m,n}^{th}\le {P_{m,n}}(k) \le P_{m,n}^A(k)\;\;\forall m,\;n,\;{k}\subseteq \mathcal{K}_d,\\
&C_2:\;{P_{m,n}}(k)=0 \;\;\forall m,\;n,\;{k}\subseteq \mathcal{K}_h,
\end{align}
\end{subequations}
where $\textbf{P}_\textbf{r}$ is the state transition probability matrix. In general, this optimization problem is impossible to be solved independently in each time slot because of the causality constraints on the variables. Then we propose to solve it by using dynamic programming.

To solve this dynamic programming problem, Bellman's equations are introduced\cite{Ad3_bertsekas1995dynamic}. Then it is expressed as the backward recursive equations, which starts from $k=K$ to $k=1$.

For $k=K$,
\begin{align}\label{bellman1}
V_K(S_{m,n}(K)))=\mathop {\max }\limits_{\pi_K } \;R_{m,n}(K),
\end{align}
and for $k={K-1,...,1}$,
\begin{align}\label{bellman2}
\begin{array}{l}
V_k(S_{m,n}(k)))\\
=\mathop {\max }\limits_{\pi_k } \; R_{m,n}(k)+{\bar V_{k + 1}}({S_{m,n}}(k),\; P_{m,n}(k)),
\end{array}
\end{align}
where the second phase consists of the future reward information from time slot $k+1$ to $K$, which is presented as
\begin{align}\label{bellman3}
\begin{array}{l}
{\bar V_{k + 1}}({S_{m,n}}(k),\; P_{m,n}(k))\\
= {{\mathbb E}_{{S_{m,n}}(k)}}\{ {V_{k + 1}}({S_{m,n}}(k+1))|\mathbb A,\;{S_{m,n}}(k)\},
\end{array}
\end{align}
where ${\mathbb E}_{{S_{m,n}}(k)}$ presents the statistical expectation of all the possible states in future time slot $k+1$ giving the current state $S_{m,n}(k)$ and the transition probabilities. (\ref{bellman3}) is considered as the optimal network throughput obtained from the future time slots. Then we maximize the cumulative network throughput from the current time slot to the last time slot resulted from the current state and the current policy.

In order to solve this problem, Bellman's equations and backward induction are used. An MDP-based power allocation algorithm is developed for each IoT user allocated to the same channel. In Algorithm \ref{optimal_online}, the initialization phase initializes the number of users allocated to the same channel according to ECAA. Then the MDP-based power allocation algorithm is performed, which includes two steps: planning step and transmission step. In the planning step, a look up table is built up to record the optimal policy $\pi^*$, that is, the optimal sequence of transmit power from time slot $K$ to 1 over all the possible states ($P_{m,n}^A(k), \;P_{m,n}^H(k-1), \;g_{m,n}(k)$). By using backward induction method, it starts from the final time slot, in which all the available power in the battery should be used, i.e., $P_{m,n}(K)=P_{m,n}^A(K)$. Then (\ref{bellman1}) is solved for all the possible values of $P_{m,n}^A(K), \;g_{m,n}(K)$. Afterwards, for $k=\{K-1,...,2,1\}$, the (\ref{bellman2}) is calculated recursively for all the possible states.
In the transmission phase, the current state ${S_{m,n}}(k)$ is known at time slot $k$, then we compare the available battery power with the pre-defined threshold to determine if the data transmission is enabled in this time slot. As a result, the optimal policy $\pi^*$ during the time frame is obtained, that is, each user knows the optimal transmit power at each time slot.

\section{Simulation}
In this section, our proposed ECAA is verified by comparing it with the baseline approaches. After using ECAA, performance of our proposed MDP-based power allocation algorithm is shown with different number of time slots. LoRa network is illustrated to demonstrate the proposed algorithms.

In the simulation, the active LoRa users are randomly distributed around the gateway, and they are located in a circle with its radius $r=1$~km. We assume LoRa technology works at 868~MHz that contains eight channels with each channel supporting different transmission rates, and the bandwidth of each channel is set as the same $B_m=125$ KHz. Each channel allows at most $D=6$ users with different SFs ranging from 7 to 12.
We set the noise power as ${\sigma ^2} =  - 174 + 10{\log _{10}}\left( {B_m} \right)$ in dBm level. The path loss exponent is set as $\alpha=3.5$ for all the communication links.
From Algorithm 3, the maximum battery capacity is $P_{max}^A=30dBm$, and the pre-defined threshold to enable data transmission is $P_{m,n}^{th}=12dBm$. A three-state Markov chain is consider to model the channel gain of the uplink transmission between LoRa users and the LoRa gateway, that is, the channels have three possible values: ``good'', ``normal'' and ``bad''. Then the channel gain set is $g_{m,n}=\{0.5\times10^{-4},\;1\times10^{-4},\;1.5\times10^{-4}\}$, with the transition matrix \cite{AD4_zhang1999finite} given by
\begin{equation}
\textbf {P}_\textbf {r}^g={
\left[ \begin{array}{ccc}
0.3 & 0.7 & 0\\
0.25 & 0.5 & 0.25\\
0 & 0.7 & 0.3
\end{array}
\right ]}
\end{equation}

For simplicity, the energy arrival process is modelled as a finite-horizon MDP with four states, the possible values are taken from $\{0, aH_e, bH_e, cH_e\}$. We set one possible value as $0$, which is the value of the harvested power of the LoRa user when the time slot $k$ is used to transmit data.
\begin{figure}[t!]
    \begin{center}
        \includegraphics[width=3.8in]{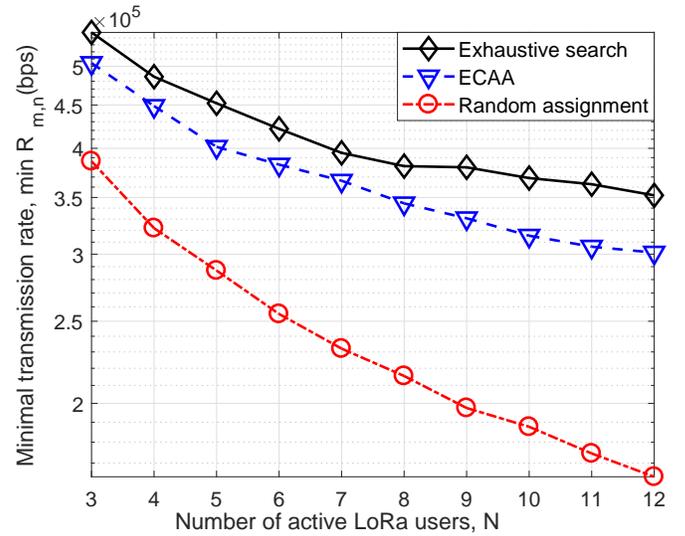}
        \caption{\small{Minimal transmission rates comparison of the three approaches, $M=3$.}}
        \label{ECAA_fixed_power}
    \end{center}
\end{figure}
\begin{figure}[t!]
    \begin{center}
        \includegraphics[width=3.8in]{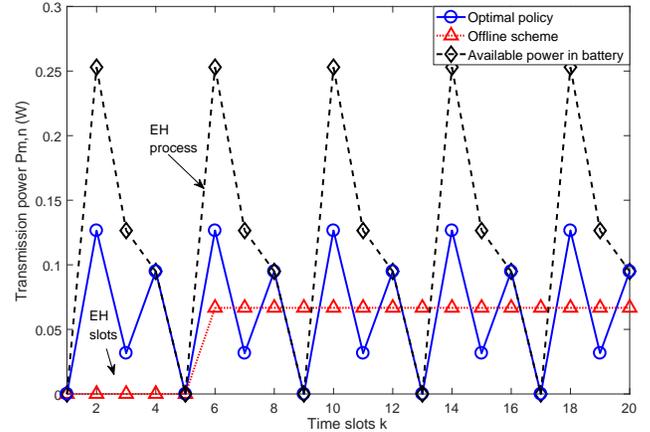}
        \caption{\small{The optimal power policy $\pi^*$, the available power in battery $P_{m,n}^A$ and the offline power allocation over time slots.}}
        \label{The optimal policy}
    \end{center}
\end{figure}
$H_e$ is the energy harvesting rate with value of $15dBm$. The transition probability matrix \cite{AD5_mao2014joint} is given as
\begin{equation}
\begin{split}
\textbf {P}_\textbf {r}^h
&={
\left[ \begin{array}{cccc}
P_{H_1H_1} & P_{H_1H_2} & P_{H_1H_3} & P_{H_1H_4}\\
P_{H_2H_1} & P_{H_2H_2} & P_{H_2H_3} & P_{H_2H_4}\\
P_{H_3H_1} & P_{H_3H_2} & P_{H_3H_3} & P_{H_3H_4}\\
P_{H_4H_1} & P_{H_4H_2} & P_{H_4H_3} & P_{H_4H_4} \\
\end{array}
\right ]}\\
&={
\left[ \begin{array}{cccc}
0.3 & 0.7 & 0&0\\
0.25 & 0.5 & 0.25&0\\
0 & 0.25 & 0.5 & 0.25\\
0 & 0 & 0.7 & 0.3 \\
\end{array}
\right ]}
\end{split}
\end{equation}
where $P_{H_iH_j},i,j\in\{1,2,3,4\}$ indicates the transition probability of the harvested energy state from state $H_i$ to state $H_j$. Then the steady state probability is given by $[P_{H_1},P_{H_2},P_{H_3},P_{H_4}]=[0.13, 0.37, 0.37, 0.13]$.

In Fig.~\ref{ECAA_fixed_power}, the feasibility of our proposed ECAA is validated with different numbers of active LoRa users. For comparison, we present another two baseline approaches as well, including random channel assignment and brute force exhaustive-search.
In this case, each LoRa user has fixed transmit power that is set as $P_{\max}^A$, which can remove the effects of different transmit powers. Compared to the brute force exhaustive-search approach, we can observe that our proposed algorithm, ECAA, achieves $90\%$ of the optimal performance. However, ECAA has much lower computational complexity as illustrated in Theorem 3.

\begin{figure}[t!]
    \begin{center}
        \includegraphics[width=3.8in]{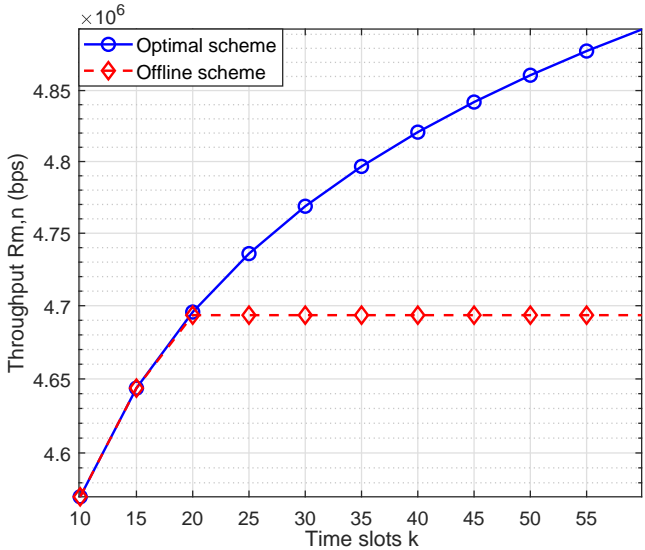}
        \caption{\small{Performance comparison of the proposed MDP-based power allocation algorithm and offline scheme.}}
        \label{throughput comparison}
    \end{center}
\end{figure}
\begin{figure}
    \begin{center}
        \includegraphics[width=3.8in]{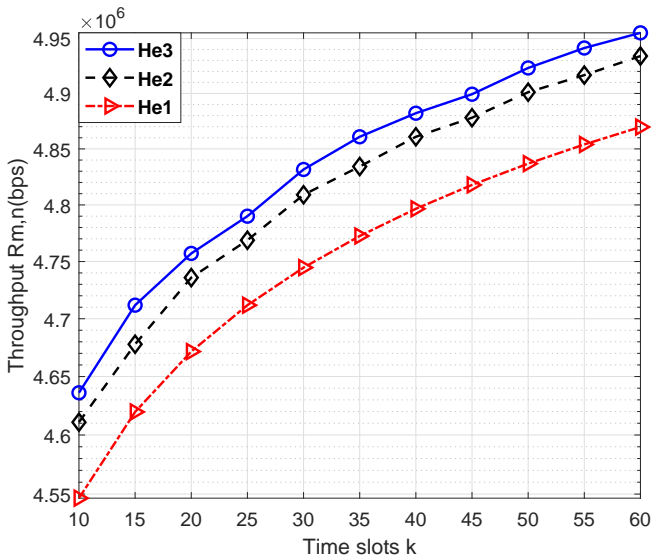}
        \caption{\small{Throughput comparison under different energy harvesting rate value vectors $\mathbf {H_e}$. }}
        \label{throughput he}
    \end{center}
\end{figure}

\begin{figure}[t]
    \begin{center}
        \includegraphics[width=3.8in]{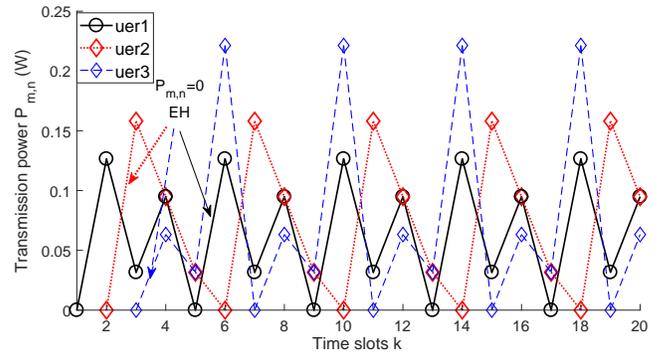}
        \caption{\small{Optimal policies for different users allocated to the same channel.}}
        \label{optimal_users}
    \end{center}
\end{figure}
\begin{figure}[t]
    \raggedleft
        \includegraphics[width=3.8in]{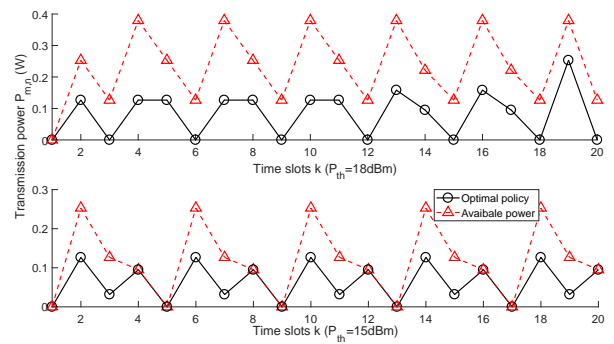}
        \caption{\small{Optimal policies under different pre-defined threshold values. }}
        \label{optimal_threshold}
\end{figure}

Fig.~\ref{The optimal policy} shows the optimal policy over the time frame obtained by the MDP-based power allocation algorithm. The energy harvesting rate value vector is set as $\{0, 2H_e, 5H_e, 8H_e\}$. The optimal policy for each LoRa user presents the optimal transmit power in each time slot, moreover, LoRa user switches to energy harvesting mode when the transmit power $P_{m,n}=0$. The offline scheme is presented for a comparison. In the offline scheme, the LoRa user harvests energy in the first few time slots, then uses all the energy stored in the battery to transmit data in the remaining time slots. As a comparison, the number of time slots used for energy harvesting is set the same as the proposed optimal scheme as shown in the Fig.~\ref{The optimal policy}. We can see that the dynamic transmit power allocation is implemented in the time frame, which is achieved by updating the available battery power in real time through the proposed algorithm.

Fig.~\ref{throughput comparison} shows the performance comparison of the proposed optimal scheme and the offline scheme in terms of achieved throughput of the whole network. we can see that the network throughput of the optimal scheme is increasing with increasing the number of time slots in the frame. However, the network throughput of the offline scheme does not increase after the number of the time slots reached 20, it remains the same. This is because that the battery capacity is limited, that is, the maximum amount of harvested power cannot exceed the battery capacity, i.e., the total available power for the rest time slots are equal to battery capacity. In Fig.~\ref{throughput he}, we set the three energy harvesting value vectors as: $\mathbf{H_{e1}}=[0, 5H_e, 8H_e,11H_e],\;\mathbf{H_{e2}}=[0, 4H_e, 7H_e, 10H_e],\; \mathbf{H_{e3}}=[0, 2H_e, 5H_e, 8H_e]$. We can see that the network throughput increases with the increasing of the energy harvesting rate value vectors.

Fig.~\ref{optimal_users} illustrates the optimal policies for different LoRa users allocated to the same channel. We use different energy harvesting value vectors to distinguish different LoRa users while their channel gains are taken from the same channel gain value vector. In Fig.~\ref{optimal_users}, energy harvesting value vectors for different LoRa users are set as $user_1=[0, 1H_e, 4H_e,7H_e],\; user_2=[0, 2H_e, 5H_e, 8H_e],\; user_3=[0, 3H_e, 6H_e, 9H_e]$, respectively. Note that there are at most 6 users transmitting data at the same time as the SF ranges from 7 to 12. In this paper, by enabling LoRa users to execute their optimal policies in different time slots, we can make each channel contain more users. This is achieved by letting the extra users to harvest energy which will not cause interferences to the users that are transmitting data.

In Fig.~\ref{optimal_threshold}, it shows that the optimal power policies with different predefined threshold values are obtained. We can observe that the battery will not run out of its power with increasing threshold. This means that the rechargeable battery do not have to discharge all of its power frequently, which will help extend its lifespan.
Moreover, the maximum transmit power will increase as shown at the bottom figure of Fig.~\ref{optimal_threshold} after $13^{th}$ time slot. This is because the user starts to transmit data with more available battery power.

\vspace{-0.05cm}
\section{Conclusions}
In this paper, the resource allocation problem for uplink communication in wireless powered IoT networks has been investigated to extend the network lifespan and maximize the network throughput of each user. At first, we have proposed an efficient channel allocation algorithm with low complexity, called ECAA, to group IoT users into available channels. Then, for users allocated to the same channel, an MDP-based power allocation algorithm has been proposed to achieve the optimal network throughput of each user. Simulations have demonstrated that our proposed ECAA achieves 90\% of the optimal performance and better performance than the random channel assignment, but with much lower complexity. The MDP-based algorithm has achieved dynamic power allocation for each user by maximizing the accumulative network throughput over a finite time frame, which has been proved to outperform the offline scheme with the number of time slots increasing in the frame. Therefore, our proposed resource allocation method has achieved a good performance considering system performance and computational complexity.

The future work will focus on intelligent data transmission in the uplink for IoT networks serving a massive number of users. In the IoT network, both the gateway and each IoT user have finite resources, such as spectrum resource, computation resource and power resource. We will investigate intelligent strategies to optimize data transmission by using limited resource through machine learning techniques. For instance, the MDP-based power allocation in this paper is solved by dynamic programming which needs known model information. Thus, exploring the model-free techniques to solve MDP problems is one of our future research directions.

\numberwithin{equation}{section}

\section*{Appendix~A: Proof of Theorem 1} \label{Appendix:B}
\renewcommand{\theequation}{B.\arabic{equation}}
\setcounter{equation}{0}
For our proposed ECAA, if the final matching $\Phi $ contains at least one more swap-blocking pair, the utility of at least one player can be reformative and the utilities of other arbitrary players wouldn't be reduced. Nevertheless, if any pair of players blocks the current matching, $\Phi $, the ECAA won't stop, that is, $\Phi $ cannot be considered as the final matching, so it will cause conflicts between players. As a result, the final matching $\Phi $ is defined as \textbf{\emph{2ES}}.

\section*{Appendix~B: Proof of Theorem 2} \label{Appendix:C}
\renewcommand{\theequation}{C.\arabic{equation}}
\setcounter{equation}{0}

For our proposed ECAA, it performs finite swap operations since there are limited number of players and each channel is restricted to accept finite number of users.
Furthermore, each swap operation will make the achieved minimal transmission rate of each channel increase. The stop condition of the swap operations is defined as when the transmission rate is satisfied in the worst case. This is because there is an upper bound for the achievable transmission rate of each channel since the spectrum resources is limited. As a result, our proposed ECAA is converged to the stable state after performing swap operations a finite number of times.

\section*{Appendix~C: Proof of Theorem 3} \label{Appendix:D}
\renewcommand{\theequation}{D.\arabic{equation}}
\setcounter{equation}{0}
From Algorithm 1 and 2, the ECAA is proposed including the initial matching as well as the swap operations, so its computational complexity is made up of two parts. In the first part, we consider the worse case that all the users are proposed to all the available channels. In this case, the computational complexity is calculated by ${\rm O}\left(MN\right)$.

In the second step, both the number of iterations in the algorithm, ECAA, and the swap operations in each iteration are contributed to the computational complexity. Nevertheless, it takes finite iterations, $I$, to reach the final matching \textbf{\emph{2ES}} though there are no closed-form expressions, which is proved in Theorem~\ref{theorem2}.
In each iteration, given any user $U_n$, $M-1$ possible swap-blocking pairs are generated since the gateway has $M$ channels to access and each user only occupies one channel at most. There are at most $D$ users are allocated to the selected channel $L_m$. Thus, there are at most $D\left( {M - 1} \right)$ possible combinations of a swap matching $\Phi$ giving any user $U_n$. In our proposed ECAA, we need to consider at most $\frac{1}{2}DN\left( {M - 1} \right)$ swap matchings in each iteration. As a result, the upper bound of the complexity in the second step is denoted by ${\rm O}\left( {\frac{1}{2}IDN\left( {M - 1} \right)} \right)$.

In  conclusion, the upper bound of the computational complexity of our proposed ECAA is calculated as ${\rm O}\left( MN+{\frac{1}{2}IDN\left( {M - 1} \right)} \right)$.

\bibliographystyle{IEEEtran}
\bibliography{IEEEabrv,mybib1}

\end{document}